\documentclass[11pt,twoside]{article}

\usepackage{asp2006}
\usepackage{epsf}
\usepackage{psfig}
\usepackage{lscape}

\begin{document}
\title{What Solar Oscillation Tell us About the Solar Minimum}   
\author{Kiran Jain, S. C. Tripathy, O. Burtseva, I. Gonz{\'a}lez Hern{\'a}ndez, F. Hill, R. Howe, 
S. Kholikov, R. Komm and J. Leibacher}   
\affil{Global Oscillation Network Group, National Solar Observatory, Tucson, AZ 85719, USA}    

\begin{abstract} 
 The availability
of continuous helioseismic data for two consecutive solar minima has
provided a unique opportunity to study the changes in the solar interior that
might have led to this unusual minimum. We present preliminary analysis of intermediate-degree mode frequencies in the 3 mHz band  during the current
 period of minimal solar activity and show that the mode frequencies 
are significantly lower than  those during the previous activity minimum.
Our analysis do not show any signature of the beginning of cycle 24 till the end of 2008.
In addition, the zonal and meridional flow patterns inferred from inverting frequencies
also hint for a delayed onset of a new cycle. The estimates of travel time 
are higher than the previous minimum confirming a relatively 
weak solar activity during the current minimum. 
\end{abstract}

\section{Introduction}
The delayed onset of solar cycle 24 and the prolonged period of minimal solar 
activity have invoked lots of interest in a variety
of studies that might be useful to characterize the sun in a quiet state.  
Studies based on the helioseismic data have provided conflicting estimates of
the length of previous cycle and shown that the present minimum is indeed 
the deepest in many aspects \citep{bison09, howe09, david09,sct09b}. Since acoustic modes spend most of the time in the outer layers of the solar interior, the intermediate- and high-degree modes can be useful in interpreting the conditions in the convection zone. In this context, we investigate the response of these modes to the period of minimum activity and compare the response with the previous one.
 
\section{Oscillation Frequencies and Solar Activity during Solar Minimum }  
The analysis presented here uses $p$-mode frequencies, $\nu$,  obtained 
from the Global Oscillation Network Group (GONG) in the 3 mHz band.
 It utilizes 21 36-day non-overlapping data sets in the frequency range 2800 
$\le \nu \le$ 3200 $\mu$Hz and degree range 20 $\le \ell \le$ 100 during
 minimum phase between  solar cycles  23 and 24 from 
2006 December 23 to 2009 January 16. We also use 21 data sets during the minimum between cycles 22 and 23 spanning over the period from 1995 June 12 to 1997 July 6. 

\begin{figure}
\plotone{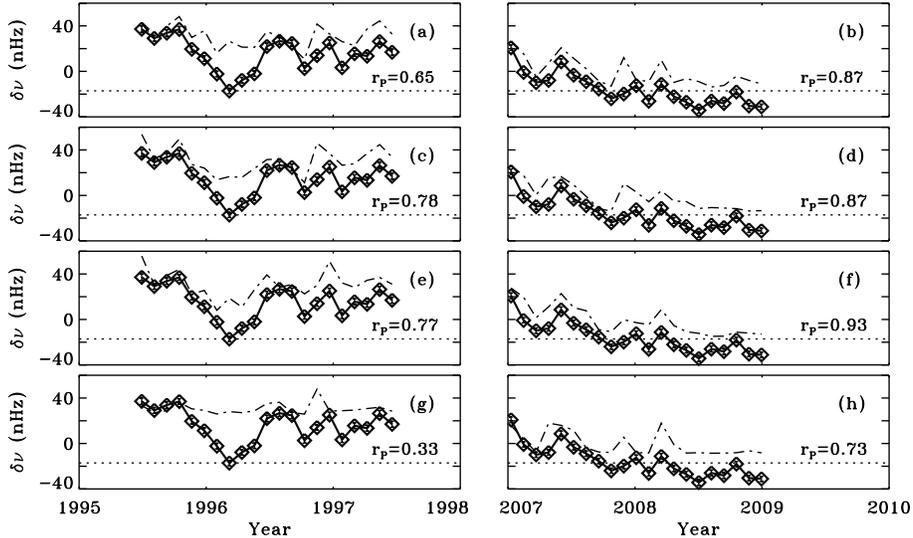}
\caption{ \label{fig:fig1} Temporal evolution of the $m$-averaged frequency shifts (symbols) with scaled
activity proxies (dashed-dot) during minima between cycles 22 and 23 (left), and 23 and 24 (right). The activity proxies used here are; (a-b) the international sunspot number, (c-d) the F$_{10.7}$ cm radio flux, (e-f) the Mt. Wilson plage strength index, and (g-h) the Mt. Wilson sunspot index. Pearson's correlation coefficients ($r{_P}$) between frequency shifts and solar activity are indicated in each panel. Dotted line represents the lowest frequency shift during the minimum between cycles 22 and 23.}
\end{figure}

The temporal variation of $m$-averaged frequency shifts ($\delta\nu$) with various measures
 of solar activity is shown in Figure~1. It should be noted that these $m$-averaged frequencies for
 global modes are averaged over all latitudes and can not be used to study the 
latitudinal distribution. The $\delta\nu$ for all 105 common 
modes are calculated with respect to the reference frequency which is 
determined by taking an average of the frequencies of a particular multiplet
 ($n,\ell$).  Although the
 frequency shifts follow the general trend of the solar activity in all
 cases, we find significantly different Pearson's correlation coefficients 
for all four activity indices (see Figure~1). These indices represent 
the changes in magnetic structures at different layers in the solar atmosphere.
 In a detailed study for the complete solar cycle, \citet{jain09} have shown that the 
correlation between frequency shifts and activity proxies  
also differs when the cycle is divided into different phases. 

It is evident from Figure~1 that the frequencies are lower and the correlation
coefficients are higher for all activity proxies during the current minimum as
 compared to the previous minimum. The best correlation is found for the 
proxy depicting the change in weak
 component of the magnetic field, represented here by the strength
 of plages (Figure~1 e-f). The other indices (sunspot number, radio flux and 
the strength of sunspots) have major contribution from the strong component. The weakest correlation is found for the field strength of sunspots which is only influenced by strong fields. These findings are in agreement with an earlier study of local modes where inclusion of the weak component of the magnetic field was found to be necessary to explain the frequency shifts during the minimal-activity phase of the solar cycle \citep{sct09a}.

Although frequency shifts are known to vary in phase with the solar activity,
we find an unusual trend during the current minimum. Solar activity shows an
 upward trend in the last quarter of 2008 while the shifts 
continue to decrease for the period considered in this analysis (for example Figure~1d)
 This anti-correlation was not seen in the minimum between cycles 22 and 23. 
\citet{david09} have interpreted the anti-correlation between low-degree frequency shifts 
from GOLF, and 10.7 cm radio flux as a signature of the onset of the new solar cycle. 
  However, their analysis suggests
 the onset of cycle 24 at the end of 2007 which is not supported by our
 analysis of intermediate-degree modes.
\begin{figure}
\plotone{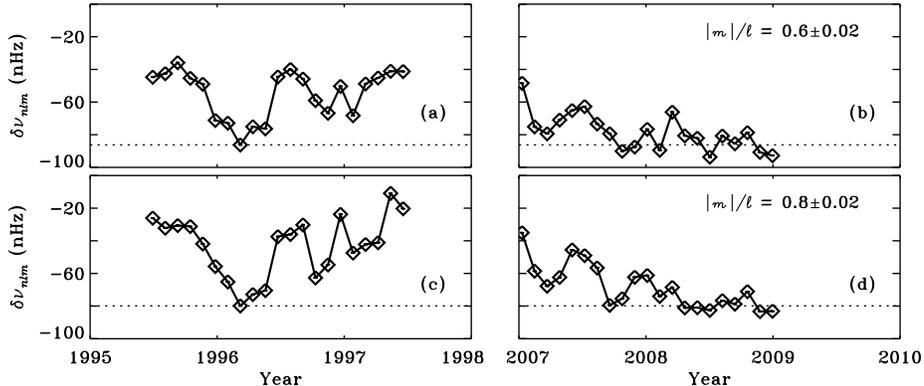}
\caption{ \label{fig:fig2} The mean variation in
frequency shifts at (a-b) $|m|/\ell\approx$ 0.6 and (c-d) $|m|/\ell\approx$ 0.8 during minima between cycles 22 and 23 (left), and 23 and 24 (right). The minimum shift at previous solar minimum is shown by the dotted line. }
\end{figure}

Since the magnetic activity related to a new solar cycle emerges first at 
mid-latitudes,  it is possible to follow the changes in oscillation modes as 
a function of the latitude using different values of $|m|/\ell$. For $|m|/\ell$=1, 
the modes are sensitive to the regions near equator while $|m|/\ell$=0
 represents modes at higher-latitudes. Figure~2 shows the mean variation in 
frequency shifts, $\delta\nu_{nlm}$, at selected values of $|m|/\ell$  representing the changes at mid/active-latitudes. As seen earlier, the frequencies are lower during
current minimum and continue to decrease with time. 
The lowest value in the left panels 
coincides with the minimum between cycles 22 and 23 providing a reasonable estimate of the 
solar minimum. Although we do not see an increase in frequencies yet, the shifts for last two data points
in Figures~2b and 2d are comparable. It might be an indication that the activity minimum has been reached in late 2008,  however, the addition of more data sets is crucial to unveil the clear picture of the onset of solar cycle 24. It is worth mentioning that the minimum seen in intermediate-degree mode frequencies will be much later than the late 2007 as reported  by \citet{david09}. 

\section{Helioseismic Inferences and Comparison with Previous Minimum} 
The oscillation frequencies can be further used to infer characteristics of the solar interior. The migrating zonal flow pattern, known as the torsional oscillation, is obtained by inverting the global mode frequencies \citep{howe09}. The analysis shows that the flow band associated with the new cycle has been found moving more slowly toward the equator than that observed during the previous minimum and suggests the length of solar cycle 23 to be approximately 12 years.

By using local helioseismology techniques, we can also study the meridional 
circulation in the solar interior. This extended solar minimum has given us 
the unprecedented opportunity to study the evolution of subsurface meridional 
flows without contamination from surface magnetic activity. A preliminary analysis of 
continuous GONG data using the ring-diagram technique suggests the formation of 
meridional flow ''bumps" under the solar surface at high latitudes before the 
surface activity appears. Detailed results will be presented in an upcoming
paper \citep{igh09}.

The analysis based on the technique of time-distance shows that the travel time
of acoustic waves varies during the solar cycle and it
 decreases due to the presence of active regions. It is seen that there are no
 significant changes at high latitudes, while travel times at low latitudes 
are decreased by about 2 seconds during the maximum activity phase. The travel times during 
2006-2008 are estimated to be 1-2 seconds higher than the previous cycle. 
 These estimates also confirms that solar activity is relatively weak during the current minimum.

In summary, we find that the onset of solar cycle 24 is delayed and we do not
see signs of the beginning of new cycle till the end of 2008. 

\acknowledgements 
This work utilizes data obtained by the Global Oscillation Network
Group (GONG) project, managed by the National Solar Observatory, which
is operated by AURA, Inc. under a cooperative agreement with the
National Science Foundation. The data were acquired by instruments
operated by the Big Bear Solar Observatory, High Altitude Observatory,
Learmonth Solar Observatory, Udaipur Solar Observatory, Instituto de
Astrof\'{\i}sico de Canarias, and Cerro Tololo Interamerican
Observatory. This study also includes data from the synoptic program at 
the 150-Foot Solar Tower of the Mt. Wilson Observatory, operated by 
UCLA, with funding from NASA, ONR and NSF, under agreement with the 
Mt. Wilson Institute. This work was supported by NASA grant NNG08EI54I.

\end{document}